# Design of a Time-to-Digital Converter ASIC and a mini-DAQ system for the Phase-2 Upgrade of the ATLAS Monitored Drift Tube detector

Yuxiang Guo[a,b], Jinhong Wang[b], Yu Liang[a], Xiong Xiao[b], Xueye Hu[b], Qi An[a], J. W. Chapman[b], Tiesheng Dai[b], Lei Zhao[a,*], Zhengguo Zhao[a], Bing Zhou[b], Junjie Zhu[b,*]

[a]*State Key Laboratory of Particle Detection and Electronics, University of Science and Technology of China, Hefei 230026, China*

[b]*Department of Physics, University of Michigan, Ann Arbor, MI, 48109, USA*

ABSTRACT

We present the second prototype of a time-to-digital (TDC) ASIC for the upgrade of the ATLAS Monitored Drift Tube (MDT) detector for High-Luminosity LHC operations. Compared to the first prototype, triple modular redundancy has been implemented for the configuration and flow control logic. The total power consumption is increased by less than 10 mW while achieving the same time resolution and channel uniformity. A mini-DAQ system has been built to verify the front-end electronics chain with the new prototype together with other ASICs and boards in triggered mode. Cosmic ray tests with a small-diameter MDT chamber indicate that the configuration and data transmission of the readout electronics perform well. It is expected that this prototype design will be used in the final production.

## 1. Introduction

The ATLAS muon spectrometer [1, 2] identifies and measures the trajectory of muons produced in proton-proton collisions at the Large Hadron Collider (LHC). It is composed of three stations (inner, middle and outer) in both barrel ($|\eta|<1.05$)** and endcap ($1.05<|\eta|<2.7$) regions. Monitored Drift Tubes (MDT) are used as precision-tracking chambers and cover the pseudorapidity range of $|\eta|<2.7$, except the endcap inner station where Cathode Strip Chambers are used in the region $2<|\eta|<2.7$ due to their higher rate capability and time resolution. MDT can determine the transverse momentum of the moun with a resolution of ~10% at 1 TeV. Resistive Plate Chambers (RPC) and Thin Gap Chambers (TGC) are used as fast trigger chambers in the barrel and endcap regions, respectively.

To benefit from the planned High-Luminosity LHC (HL-LHC) upgrade, ATLAS plans to have an accept rate and a latency at the first trigger level of 1 MHz and 10 us, respectively [3]. A maximum hit rate of 400 kHz per tube is expected. In addition, the MDT detector will be used at the first-level trigger to improve the muon transverse momentum resolution and reduce the overall muon trigger rate [4]. A new MDT trigger and readout system has been proposed and is currently under development.

* Corresponding authors.
E-mail address: junjie@umich.edu (Junjie Zhu) and zlei@ustc.edu.cn (Lei Zhao).

**ATLAS uses a right-handed coordinate system with its origin at the nominal interaction point (IP) in the center of the detector and the z-axis along the beam pipe. The x-axis points from the IP to the center of the LHC ring, and the y-axis points upward. The pseudorapidity is defined in terms of the polar angle θ as $\eta = -\ln\tan(\theta/2)$.

The proposed MDT trigger and readout chains for HL-LHC runs are described in [5]. Following the passage of a muon through a tube, the electrons from the primary ionization clusters drift to the central wire along radial lines. The difference between the earliest arrival time of the signal at the wire and the reference time provided by RPC/TGC gives the drift time of the muon hit, and this drift time is used to determine the drift radius. The earliest arrival signal at the wire is captured by an Amplifier/Shaper/Discriminator (ASD) ASIC [6, 7]. In addition, ASD measures the pulse height of the detector signal for the first ~20 ns, which allows the monitoring of the gas gain as well as for the pulse-height-dependent slewing corrections to the timing measurement. The pulse height of the signal is encoded as the time interval between the leading and trailing edges of the ASD output pulse. A Time-to-Digital Converter (TDC) ASIC is used to perform the required time measurements on 24 channels coming from three ASD ASICs. The TDC outputs are then transmitted to a Chamber Service Module (CSM) using two serial data lines running at a line rate of 320 Mbps each. The CSM multiplexes data from up to 18 TDCs and sends the data via two optical fibers (each at a rate of 10.24 Gbps) to the MDT trigger processor. The MDT trigger processor uses the reference time provided by RPC/TGC trigger chambers to extract relevant muon hits, and applies segment-finding and track-fitting algorithms on matched hits to determine the momentum of the incoming muon. The fitted muon tracks are sent to the global ATLAS first-level trigger processor. All matched hits are sent to the ATLAS readout system for events satisfying the first-level trigger requirements.

## 2. The Time-to-digital converter ASIC

### *2.1 Introduction*

The TDC digitizes both leading and trailing edges of discriminated signals from three ASDs (24 channels in total) with a least count of 0.78 ns and a digitization range of 17 bits to cover the time of one LHC orbit cycle (~89 μs). This time digitization is the basis for all subsequent trigger and readout processing. The overall TDC block diagram is shown in Figure 1. The TDC can either record leading and trailing edges separately or pair the arrival time of the leading edge with the time difference between two edges. The TDC will run in the default triggerless mode to shift out the digitized time immediately to the CSM. In addition, a triggered mode is implemented for chamber testing and test beam studies. In this mode, time measurements are buffered waiting for the trigger accept signal which is used in a time-matching algorithm to select hits of interest for outputs. The design and performance of the first TDC prototype (TDCv1) using the TSMC 130 nm CMOS process can be found in [5]. The design used the ePLL block originated from the CERN microelectronics group [8, 9] and a high-speed differential IO design from the AGH University of Science and Technology [10]. To increase the radiation hardness of the ePLL, the voltage-controlled oscillation (VCO) in the original CERN's design was optimized to have minimal phase jumps after a single-event transient. The original design was ported to the TSMC 130 nm CMOS technology while the same VCO topology was used. This prototype has all HL-LHC related design features included except radiation-tolerance enhancement for critical components.

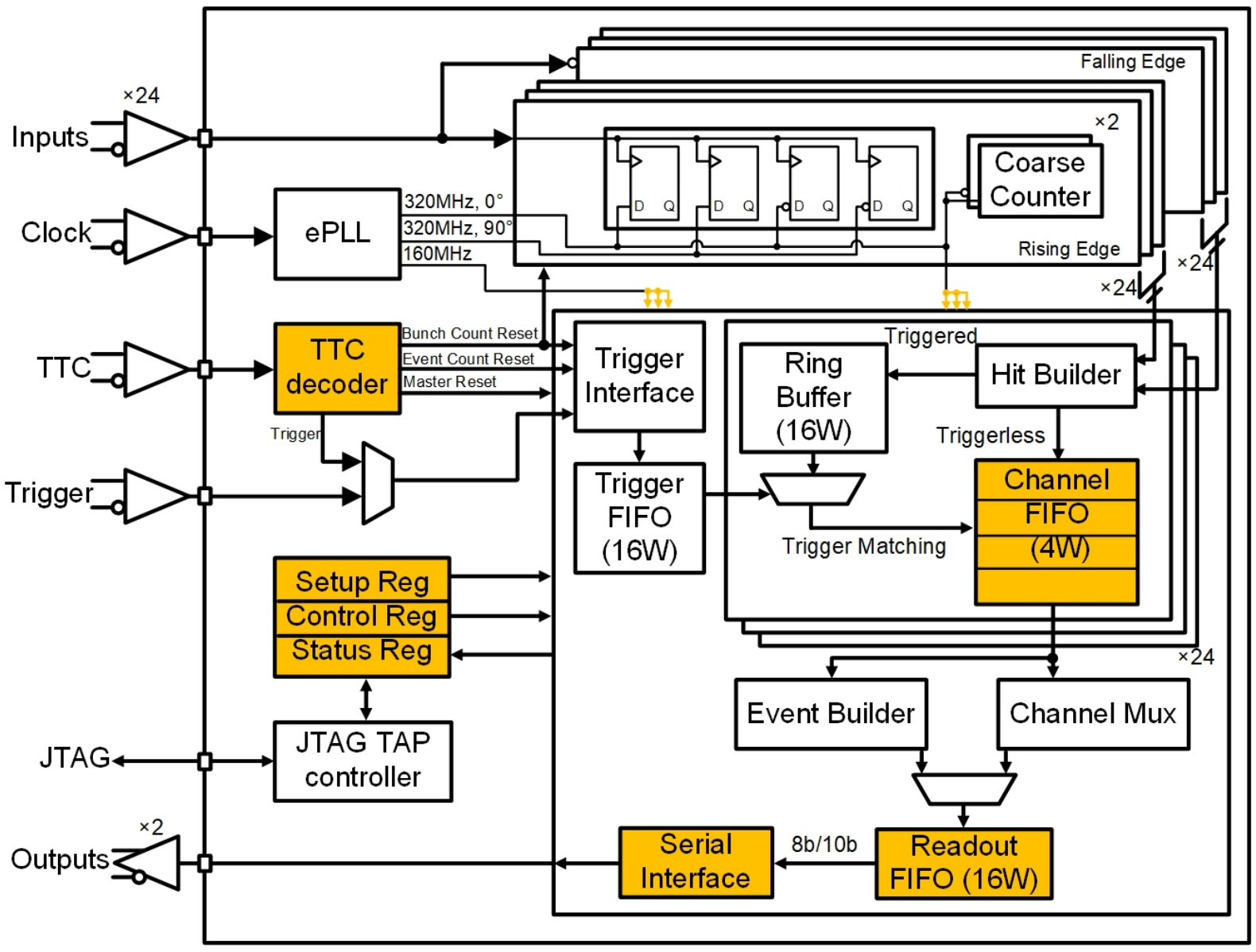

Figure 1. The block diagram of the TDC ASIC. Boxes highlighted in yellow indicate blocks that are Triple Modular Redundancy (TMR) protected.

All front-end electronics (ASD, TDC, front-end mezzanine cards, and CSM) are expected to receive significant levels of radiation during HL-LHC runs, thus it is important to have these components to be radiation-tolerant. Compared to the current LHC runs, HL-LHC will have an integrated luminosity of 4000 $fb^{-1}$ and an instantaneous luminosity of 5-7×$10^{34}$ $cm^{-2}s^{-1}$. A maximum hit rate of 400 kHz per tube is expected. A high level of reliability of the MDT electronics must be maintained during the ~15 years of operation of the experiment. The radiation in the ATLAS detector is predominately secondary particles produced by interactions of the detector elements with the primary particles produced from the collisions. There are multiple ways that radiation can affect the operation of the electronics, which normally can be divided into three categories: total ionizing dose (TID), displacement damage (non-ionizing energy loss: NIEL), and single-event effects (SEE). A radiation map of the TID in the region of the MDT electronics, the corresponding radiation map for NIEL, and the radiation map of hadrons and neutrons capable of causing SEE in electronics can be found in [4].

For the barrel inner station, the electronic boards will be located at a radial distance from the beam line (z-axis) of R = (4.4 - 4.6) m. For the endcap middle station, the boards will be located at a longitudinal position of Z = ± (13.9 - 14.3) m. The highest radiation levels for the ATLAS MDT read-out are in the BIS78 region (R = 4.4 - 4.5 m and Z = ± (5.9 - 7.3) m) based on the simulation [4]. To quantify the qualification process, a set of radiation-tolerance criteria (RTC) has been developed following a detector-wide radiation policy set by the ATLAS collaboration [11]. The criteria results from simulated radiation levels multiplied by safety factors. The simulated radiation level (SRL) dose and safety factors including simulation deviation (SFsim), low dose rate effect (SFldr) and lot-to-lot deviation (SFlot) are shown in Table 1, representing the BIS78 region with an integrated luminosity of 4000 $fb^{-1}$. It indicates that the TDC must handle a TID of 12.6 krad, an integrated 1-MeV neutron equivalent flux of $2.64\times10^{12}$ n/cm$^2$, and an integrated flux of $4.48\times10^{11}$ n/cm$^2$ for hadrons with kinetic energy above 20 MeV. Since the TDC will be manufactured in a 130 nm CMOS technology, NIEL effects are not considered here as modern CMOS integrated circuits are insensitive to displacement damage.

| | SRL | SFsim | SFldr | SFlot | RTC |
|---|---|---|---|---|---|
| TID (krad) | 5.6 | 1.5 | 1.5 | 1 | 12.6 |
| NIEL ($10^{11}$ n/cm$^2$) | 13.2 | 2 | 1 | 1 | 26.4 |
| SEE ($10^{11}$ n/cm$^2$) | 2.24 | 2 | 1 | 1 | 4.48 |

Table 1. Radiation-tolerance criteria with safety factors for the BIS78 region with an integrated luminosity of 4000 $fb^{-1}$.

### *2.2. Triple modular redundancy (TMR) for the TDCv2 ASIC*

The TDC was implemented as an ASIC instead of a components-off-the-shell (COTS) design in view of its cost, power consumption and radiation tolerance. For the 130 nm CMOS technology, performance degradation due to the cumulative effects from the specified TID (12.6 kRad) is insignificant [12]. As a result, radiation tolerance of the TDC design is mainly focused on mitigating SEE effects. SEE is a collection of immediate effects caused by the interaction of high-energy particles with electronic elements in integrated circuits. These effects can be divided into two broad categories: recoverable errors (single-event transients, single-event upsets, single-event function interrupts) and non-recoverable errors (single-event latch-up, single-event burnout, single-event gate rupture). While a power-cycle will be performed for non-recoverable errors from a global power/system monitoring, the disruption from recoverable errors will be reduced by implementing TMR logic for all critical blocks and clock trees. For the rest of the paper, we use SEU to indicate all recoverable errors.

A complete TMR implementation with every register triplicated and majority voters used to make decisions on the output would provide the best protection against SEUs, however it will significantly increase the power consumption, up to a factor of three in the worst scenario. Given the fact that there will be no cooling system planned for the upgraded MDT front-end electronics, it is crucial for the TDCv2 design to limit its maximum power consumption to be below 350 mW. Test of the TDCv1 showed an average power consumption around 250 mW, of which ~146 mW came from the TDC logic [5]. It is thus impossible to make a full TMR of the whole TDC logic and stay within the total power budget. Fortunately, a full TMR of the whole TDC logic is not necessary since a typical muon track travels through six to eight layers of tubes, errors on the drift time measurement from a single tube will not greatly affect the final track reconstruction. A partial TMR scheme is thus preferred by balancing between protection of the most critical electronics blocks and the increase of the overall power consumption.

The TDC logic is grouped into four different categories: sampling logic, memory logic, flow control logic and configuration logic. The sampling logic includes sampling registers and coarse counters for all 24 channels. Whenever a hit-ready signal is generated in the sampling logic, a TDC data word is assembled in the memory logic and then read out by the flow control logic. The TDC data word format and the readout procedure are defined by the configuration logic. An SEU occurring in different categories of the logic may affect the TDC output at different levels. As for the memory logic, an SEU can make a valid data word void; while for the sampling logic, it may either trigger a false TDC data word or corrupt the coarse counter which voids all data of this channel afterwards. Fortunately, the coarse counter is reset regularly by the global bunch count reset signal at a period of 90 μs. As for the configuration or flow control logic, an SEU possibly makes the whole chip running in an undesired mode or get stuck in a self-lock state, which requires a reconfiguration or a hard reset to restore the chip to normal operation. Therefore, we implemented TMR in the configuration and flow control logic, but not in the memory logic.

The TMR is realized in circuits by replacing registers with TMR cells for the configuration and flow control logic. A basic TMR cell consists of three registers at the input, which are clocked by three clock trees, and three majority voters at the output [13], as depicted in Figure 2(a). This design ensures that the TMR cell is immune from SEUs not only in the registers, but also in the voters, combinational logic and clock paths. Since TMR only protects against single bit errors and cannot handle multiple bit upsets, a data-scrubbing scheme is implemented in the configuration logic. Figure 2(b) shows a basic TMR cell with data scrubbing. At the end of the configuration procedure, a user-input value is passed to the registers. After that, outputs of the voters are fed back to the inputs of these registers, ensuring that any SEU will not last more than one clock cycle. No extra care of SEU accumulations is needed for the flow control logic since the registers are updated every clock cycle.

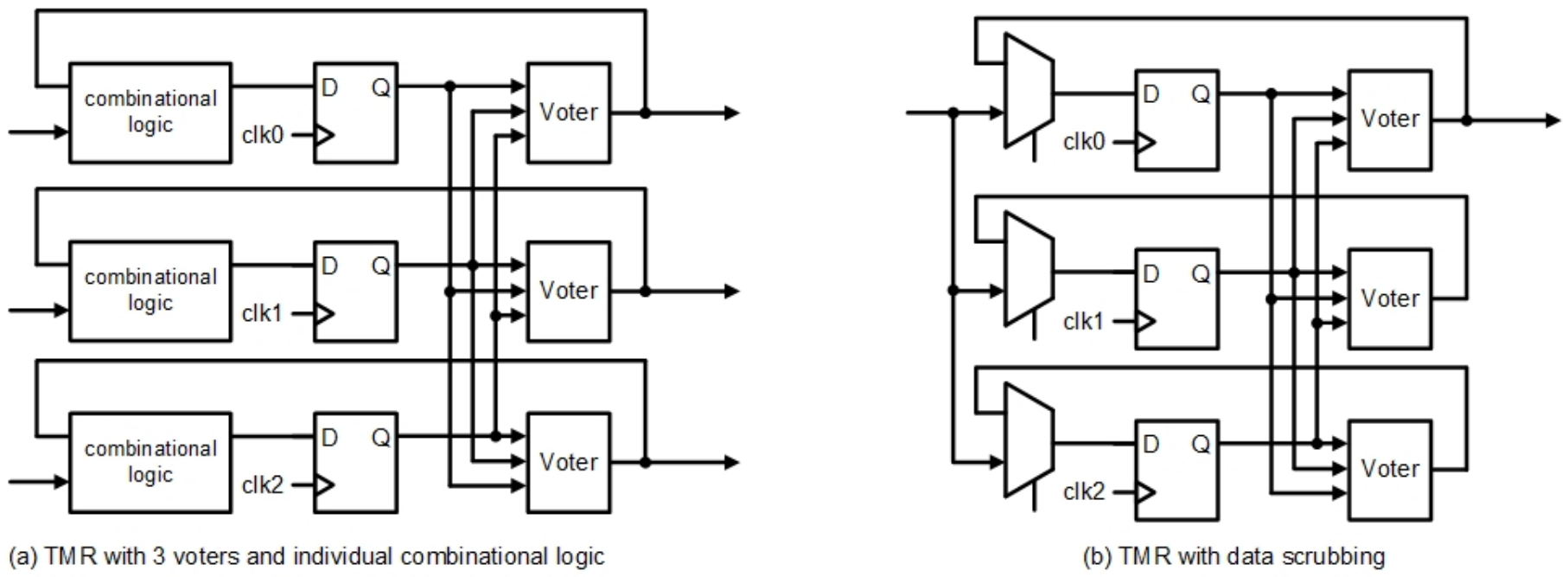

Figure 2. TMR basic cells used in TDCv2: (a) TMR with three voters and individual combinational logic (used in the flow control logic); (b) TMR with data scrubbing (used in the configuration logic).

The specific blocks that are TMR protected are shown in yellow in Figure 1. For the configuration logic, shift register chains are TMR protected with data scrubbing. For the flow control logic, the Time, Trigger and Control (TTC) block which gives trigger and reset signals, all FIFO control signals (FIFO memory excluded) and serial interface block are all TMR protected. To indicate if the circuits are interfered by an SEU, cyclic redundancy check is implemented in the configuration logic and an 8b/10b encoding scheme is used in the serial interface.

### *2.3. Performance of the TDCv2 ASIC*

The TDCv2 prototype was fabricated in late 2019 using the TSMC 130 nm CMOS process. A new 12×12 Ball Grid Array (BGA) substrate was also designed. The BGA package is used since the bonding wires are shorter than those of a QFN100 package used for the TDCv1, yielding a lower packaging failure rate.

A test board, shown in Figure 3(a), has been designed and fabricated to evaluate the performance of the TDCv2. A 144-pin BGA socket from Ironwood Electronics is utilized to hold the TDCv2 ASIC. This test board is paired with a field programmable gate array (FPGA) evaluation board, which generates pulse signals as inputs to the TDC. The TDC digitizes these pulse signals and the measured time information is then transmitted back to the FPGA board. An adapter board has been designed so that long cables can be used to isolate the FPGA board from the radiation source during irradiation tests.

Code density test results show that the uniformity of the fine-time bin size of all 48 slices (each channel has two independent TDC slices) is equivalent to the TDCv1, with a mean value of 781 ps and a variation of less than 40 ps (±5% of the bin size), as shown in Figure 3(b). The scanning of the timing performance as a function of the time interval indicates that the timing performance is close to the theoretical prediction, and contributions from other sources are negligible. All functions of the TDC have been found to work properly. The total power consumption of TDCv2 running in triggerless mode with a 400 kHz input rate per channel is 257 mW, compared to 250 mW for the TDCv1. The measured power consumption agrees well with our simulation done during the design stage. These tests indicate that the TMR implemented in the configuration and flow control logic does not significantly increase the power consumption, and does not degrade the timing performance or disrupt the TDC functions. A radiation test of TID and SEU has been delayed until 2021 due to COVID-19 restrictions.

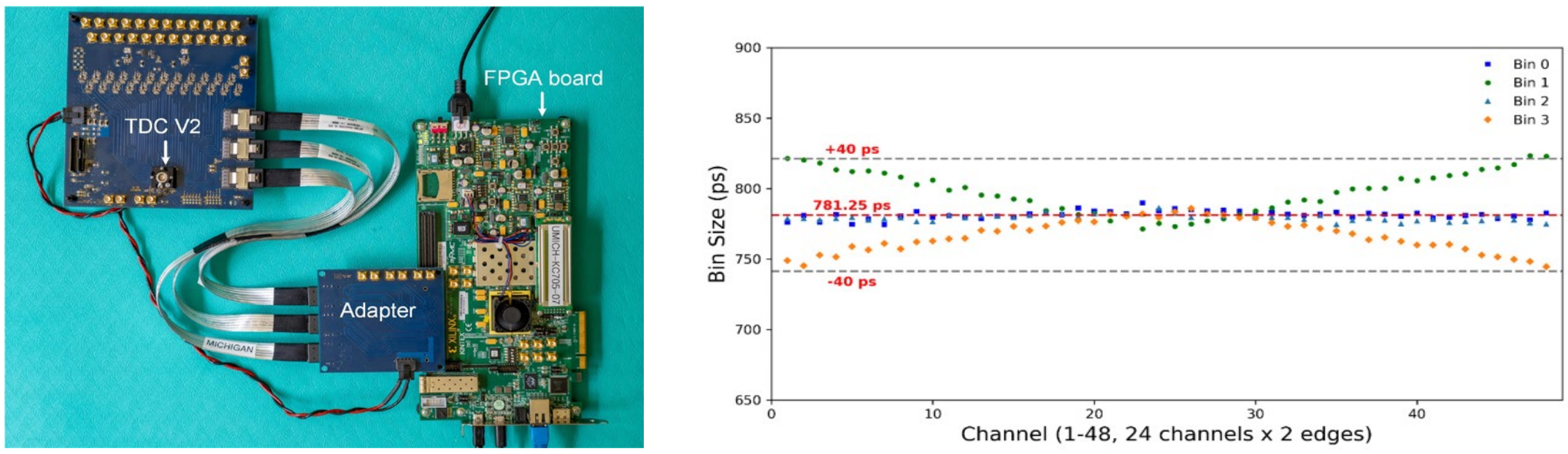

(a) (b)

Figure 3 (a) The TDCv2 test system; (b) fine-time bin size for all 48 slices.

## 3. A mini-DAQ system to test the TDCv2 ASIC

### *3.1. TDC timing Performance*

To verify the TDCv2 performance in the front-end electronic chain, a mini-data acquisition (mini-DAQ) system has been designed. This mini-DAQ system utilizes some of the current MDT readout electronics [14], such as the CSM for communications with the TDCv2, the TTCvi module for clock and trigger distribution, and the MDT-DCS Module (MDM) for initialization and configuration of the CSM and the mezzanine boards. This mini-DAQ system is used to readout a 432-channel small-diameter Muon Drift Tube (sMDT) prototype chamber. Cosmic-ray data has been taken to determine the performance of both the detector and front-end electronics.

Figure 4 illustrates the overall block diagram of the mini-DAQ system used for the cosmic ray tests in our laboratory. The sMDT prototype chamber has eight layers of drift tubes with 54 tubes per layer. The drift tubes are filled with a gas mixture of 93% Argon and 7% CO2. A high voltage of 2,730 volts is applied to the central wire. Compared to the current ATLAS MDT drift tubes, the sMDT diameter is halved to 1.5 cm and thus a higher rate capability is expected. ATLAS will use this technology to replace the small-sector MDT chambers in the barrel inner layer for HL-LHC runs [4]. A three-layer stacked mezzanine card designed by the ATLAS group at Max Planck Institute for Physics at Munich is used to cope with the limited mounting space on the sMDT chamber. This stacked card integrates three ASD ASICs and one TDCv2 ASIC, as well as input protection circuitry. It mounts directly to the signal and ground pins of the sMDT chamber, and communicates with the CSM through a 40-pin flat cable. Each card handles 24 tubes and uses the final version of ASD chips [6, 7]. The mini-DAQ system uses the legacy CSM, while the new CSM for HL-LHC runs is still under development.

Trigger and reference time information are generated by a 1m × 2m scintillator placed on the top of the chamber. In this setup, only the TDC triggered mode is used, thus the digitized time information for the discriminated signals are firstly stored inside each TDC. The signals from two photomultiplier tubes placed at both ends of the scintillator are discriminated and a coincidence trigger is formed. The resulting trigger signals are then sent to the CSM board through the TTCvi module, which also provides the global clock and synchronization signals for the whole system. The CSM distributes the trigger signals to all connected mezzanine cards. After receiving the trigger signal, matched data from up to 18 TDCs are multiplexed and reassembled by the CSM. The CSM output data is transmitted through an optical fiber to a Linux computer with a Filar card at a rate of 1.6 Gbps. The mini-DAQ system is configured via JTAG signals distributed by the MDM. A picture of this setup is shown in Figure 5.

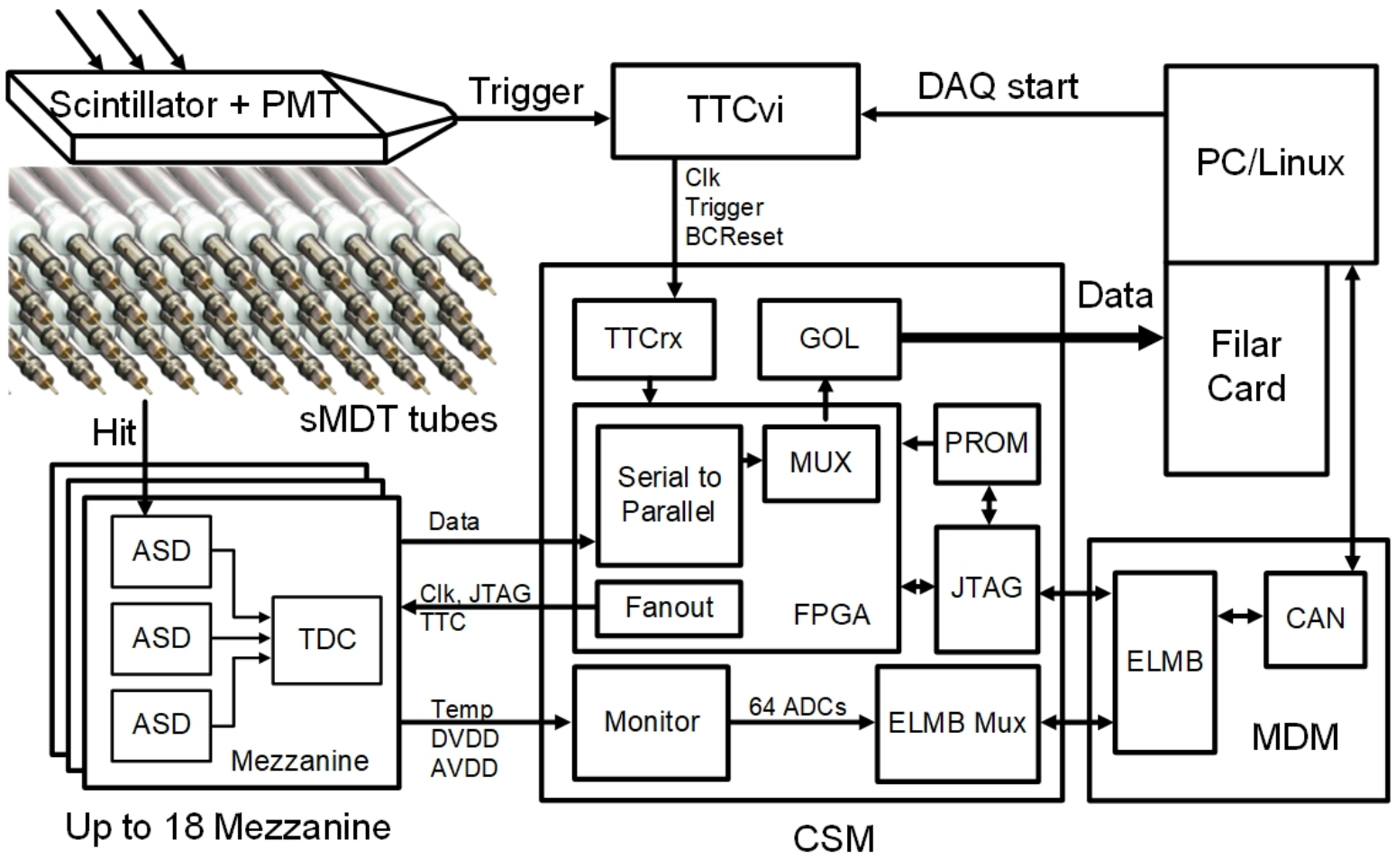

Figure 4. Block diagram of the mini-DAQ system.

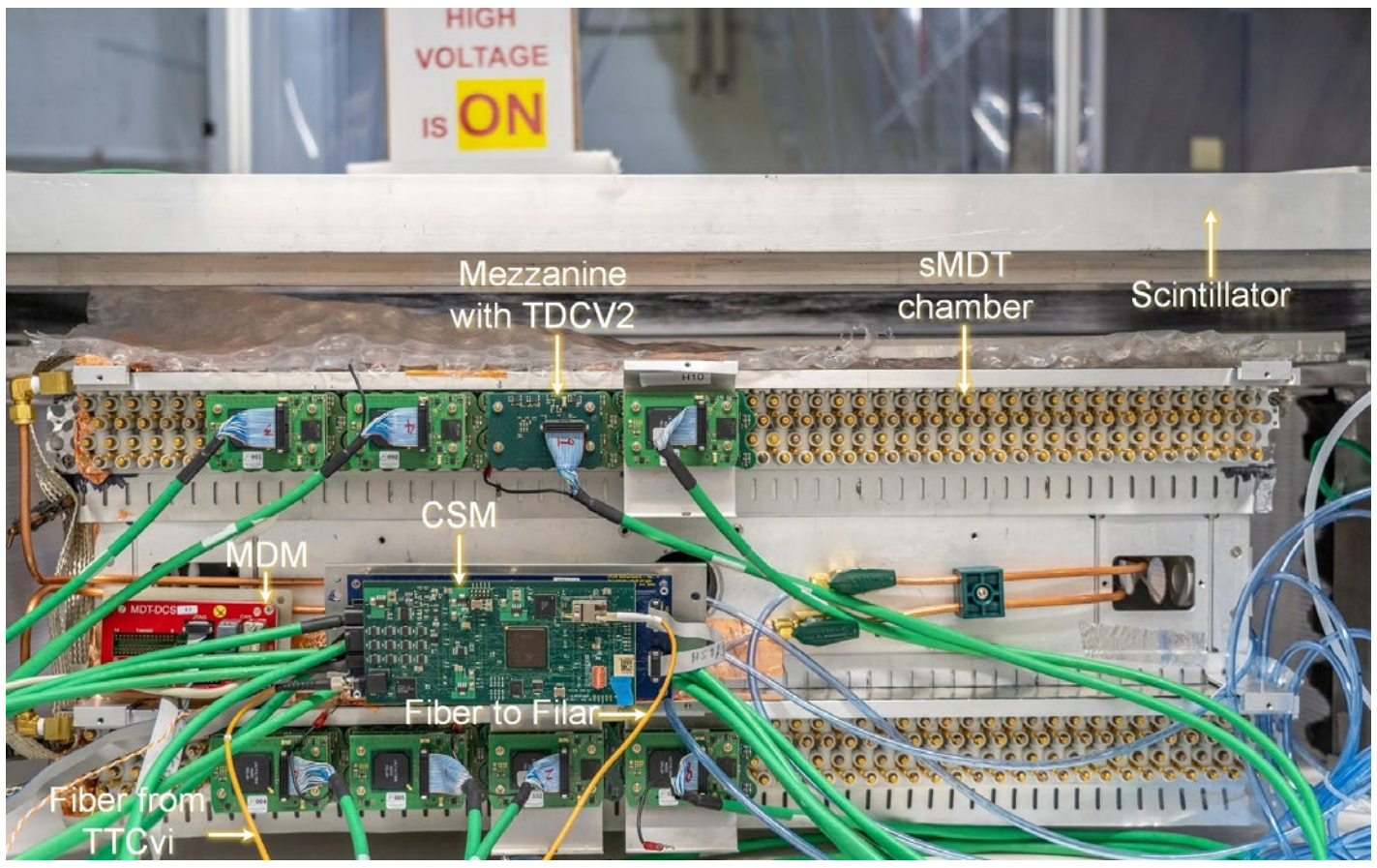

Figure 5. Setup of the cosmic ray test stand in our laboratory.

The mini-DAQ system uses a graphical user interface (GUI), shown in Figure 6, controlling configuration of the hardware, data acquisition, and monitoring. This GUI has been developed using National Instruments LabWindows. Four main functions are provided: 1) configuration of all boards and ASICs by accessing the JTAG chain of the devices on the CSM board and the TDCs and ASDs on the mezzanine cards; 2) control of the TTCvi module to distribute triggers and time alignment signals to the mezzanine cards; 3) control of the Filar card for data acquisition; 4) monitoring data/error in real time by analyzing the collected data using a Linux PC.

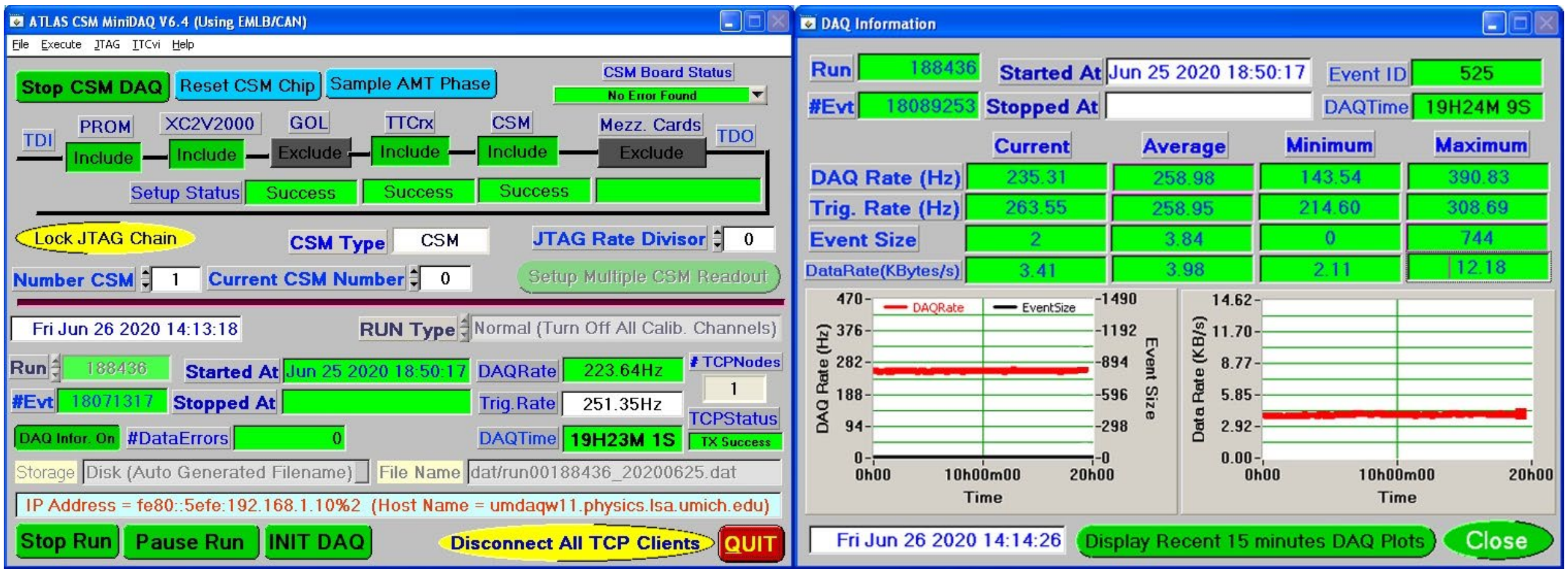

Figure 6. GUI of the mini-DAQ system to test the TDCv2 ASIC in the triggered mode. The main mini-DAQ control GUI is shown on the left and the monitoring information of the mini-DAQ is shown on the right.

Some cosmic ray test results are shown in Figure 7. Figure 7(a) shows the muon drift time spectrum, which indicates the signal earliest arrival time of drift electrons with respect to the trigger time, with a tick size of 0.78 ns. The rising edge comes from muons passing close to the tube wire, while the falling edge at the end of the spectrum comes from muons passing close to the tube walls. The maximum drift time is found to be 193 ns [15], compared to 800 ns for the legacy MDT chamber. The pulse width of the ASD output, which corresponds to the signal size, is shown in Figure 7(b). The pulse height information is used as a time slew correction to improve the resolution of the drift time measurement. These results indicate that the TDCv2 design works correctly in triggered mode.

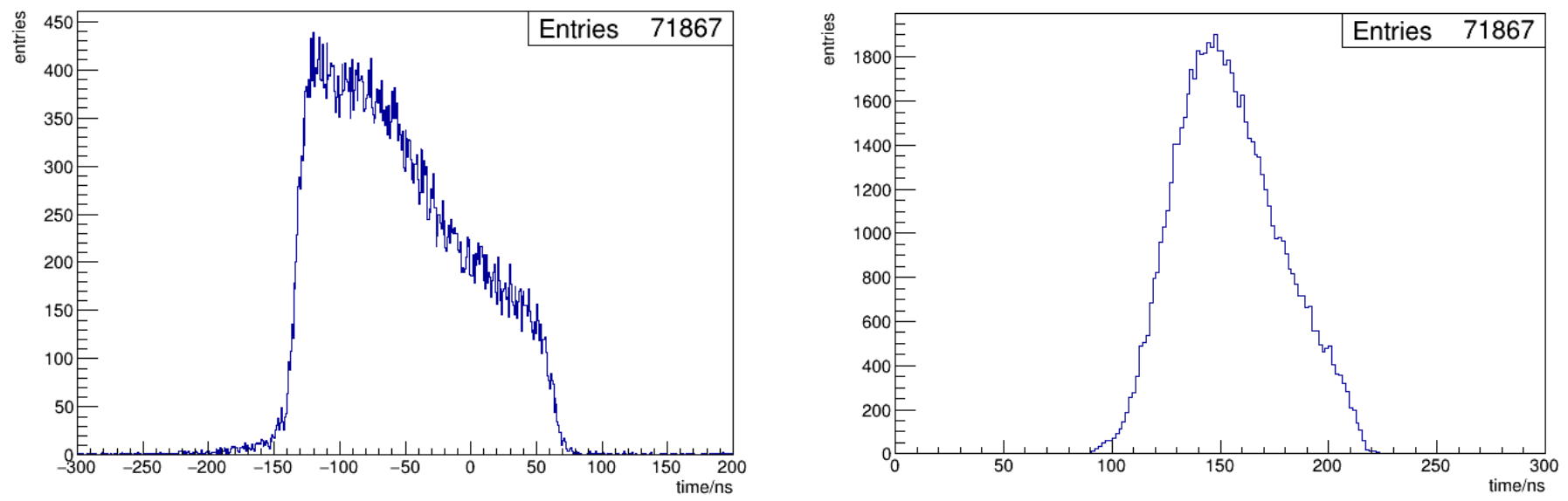

(a) (b)

Figure 7. Results from the cosmic ray tests using an sMDT prototype chamber: (a) the drift time spectrum; (b) the pulse height spectrum.

## 4. Summary

A new prototype of the TDC ASIC for the upgrade of the ATLAS MDT front-end electronics at the HL-LHC has been designed and fabricated in the TSMC 130nm CMOS technology. Triple modular redundancy has been implemented in the new design for the configuration and flow control logic. The total power consumption increased by less than 10 mW compared to the previous TDC prototype while achieving the same time resolution and channel uniformity. A mini-DAQ system has been built to verify the front-end electronics chain with the TDCv2 together with the final version of the ASD ASIC and the legacy CSM in triggered mode. Cosmic ray tests with an sMDT chamber indicate that the configuration and data transmission of the readout electronics perform well. A radiation test of the TDCv2 will be performed in 2021. It is expected that this prototype design will be used in the final production of 15,300 ASICs needed for the upgrade.

**Acknowledgments**

The USTC and UM personnel are supported by the National Key Program for S&T Research and Development (Grant NO. 2016YFA0400100) and US National Science Foundation under contracts PHY1948993, respectively. We acknowledge the help from CERN microelectronics group for providing the original ePLL design, S. Bugiel, et al. for providing high speed differential IOs, and the ATLAS group at Max Planck Institution at Munich for providing the stacked mezzanine card. The authors would also like to thank Edward Diehl from the University of Michigan for his help in this work.